\documentclass[twocolumn,prl,showpacs]{revtex4-1}

\usepackage{graphicx}

\usepackage{epsfig}

\begin{document}

\title{Formation of the Smectic-B Crystal from a Simple Monatomic Liquid }

\author{A. Metere$^1$ T. Oppelstrup$^{2}$ S. Sarman$^1$, A. Laaksonen$^1$ and M. Dzugutov$^{3}$}

\affiliation{$^1$ Dept. of Materials and Environmental Chemistry, Stockholm University, Arrhenius V\"{a}g. 16C S-106 91 Stockholm, Sweden\\ $^2$ Lawrence Livermore National Laboratory - 7000 East Avenue, Livermore, California 94551, USA\\ $^3$  Dept. of Mathematics and Centre for Parallel Computers, Royal Institute of
Technology, S-100 44 Stockholm, Sweden}

\begin{abstract}
  We report a molecular dynamics simulation demonstrating that the
  Smectic B crystalline phase (Cr-$B$), commonly observed in mesogenic
  systems of anisotropic molecules, can be formed by a system of
  identical particles interacting via a spherically symmetric
  potential. The Cr-$B$ phase forms as a result of a first order
  transition from an isotropic liquid phase upon isochoric cooling at
  appropriate number density.  Its structure, determined by the design
  of the pair potential corresponds to Cr-$B$ structure formed by
  elongated particles with the aspect ratio $1.8$. The diffraction
  pattern, and the real-space structure inspection demonstrate
  dominance of the ABC-type of axial layer stacking. This result opens
  a general possibility of producing smectic phases using isotropic
  interparticle interaction both in simulations and in colloidal
  systems.
  \end{abstract}

\date{\today}
\maketitle

Computer simulations using particles are now well-established tools
for investigating different aspects of liquid crystals \cite{care,
  bates}. The most interesting of these are smectic phases where the
molecules, besides uniaxial directional order, form layered structures
\cite{deGennes and Prost, Chandrasekhar}. In smectic-$A$ phase
positional order within layers is entirely absent, whereas hexatic
smectic-$B$ phase is characterised by a short-range hexagonal
intralayer order \cite {Doucet}. This excludes any long-range
periodicity and keeps the system fluid due to Landau-Peierls
instability \cite{landau}. Another modification of smectic-$B$ phase
is a true 3D crystal (Cr-$B$). The nature of this phase was a
controversial issue for some time, until its global 3D positional
order was established in 1979 \cite{leadbetter2}.

The simulation studies of models forming liquid-crystal phases provide
a unique way to establish a relation between the molecular-level
properties and macroscopic behaviour.  These simulations demonstrated
that the remarkable polymorphism of the mesogens forming liquid
crystal phases can be reproduced using quite simple particle
models. Systems of elongated molecules interacting via Gay-Berne (GB)
potential \cite{Zannoni, bates2, bates3, miguel} have been found
successful in reproducing the liquid-crystal phase behaviour in a
variety of simulation experiments. These models were also exploited
for calculating transport properties of the liquid crystals
\cite{sarman}. Moreover, it was proved possible to reduce the models
of anisotropically interacting GB particles to hard spherocylinders
\cite{Bolhuis, Allen}, which appears to indicate the entropic
origin of the LC phases due to geometry of excluded volume.

This experience poses a question of conceptual interest for the
statistical mechanics of condensed matter: how far further can the
particle models successfully reproducing the basic features of smectic
phases be simplified? In particular, is the anisotropy of the
constituent molecules a prerequisite to producing anisotropic
structures like those of smectic phases, and could the entropic
effects of the particle geometry be compensated by an appropriately
designed pair interaction?

This question is addressed in a molecular-dynamics simulation that we
report in this Letter. We demonstrate that the Cr-$B$ phase, a
characteristic freezing form of mesogenic systems of anisotropic
molecules, can be formed in a system composed of a single sort of
particles interacting via a spherically-symmetric potential. The
crystal occurs upon cooling as a result of a first-order phase
transition from isotropic liquid. It represents a uniaxial structure
composed of stacked layers with hexagonal close-packed intralayer
structure. The structure is consistent with the experimentally
observed Cr-$B$ structures \cite{leadbetter2,leadbetter1},
demonstrating predominantly ABCA sequence in layers stacking. This
result opens a perspective of producing other (non-crystalline) types
of smectic phases, like Smectic-A and hexatic phase. It also suggests
that this class of layered mesomorphs can possibly be produced in
systems of spherically-shaped colloidal particles.

\begin{figure} 
\includegraphics[width=6.cm]{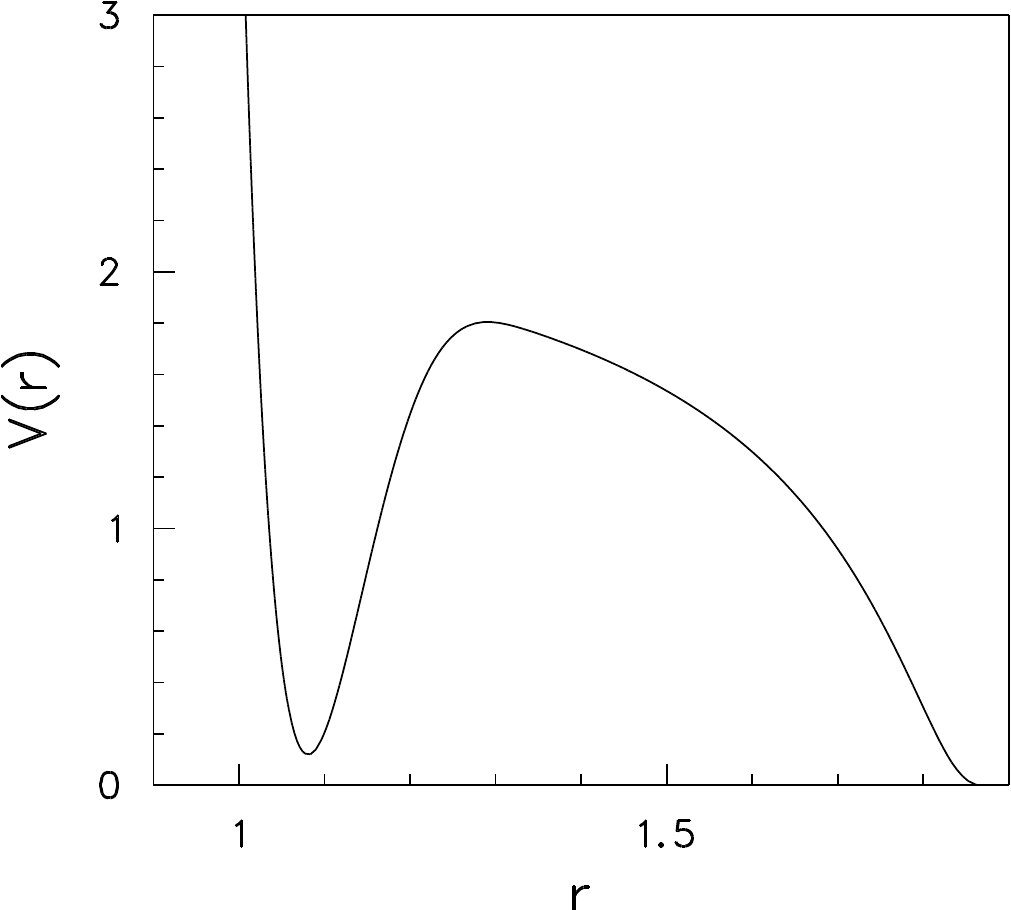}
\caption{ Pair potential}
\label{fig1}
\end{figure}

The results we report here have been produced in a
molecular-dynamics simulation of a single-component system
comprising 16384 particles. The interparticle interaction was
assumed to be spherically symmetric, described by the pair
potential presented in Figure \ref{fig1}. The functional form of the
potential energy for two particles separated by the distance $r$
is:
\begin{equation}
V(r) = a_0\left(r^{-m} - a_1\right) H(r,b_1,c_1) + H(r,b_2,c_2),
\end{equation}
where
\begin{equation}
H(r,b,c) = \left\{ \begin{array}{ll}
\exp\!\left(\frac{b}{r-c}\right) &\quad r < c \\
0 &\quad r\geq c.
\end{array} \right.
\end{equation}
The values of the parameters are presented in Table I. The first term
describes the short-range repulsion branch of the potential, and its
minimum, whereas the second term expresses the long-range
repulsion. All the thermodynamic quantities we report are expressed in
terms of the reduced units that were used in the definition of the
potential. Note that the steepness of the short-range repulsion, and
the position pf the minimum, are consistent with those in the
Lennard-Jones (LJ) potential \cite{hansen}, which makes it possible to
compare the reduced number densities of the two systems.
\begin{table}
\begin{tabular}{cccccccc}
 
\hline 
\hline 

m & $a_0$ & $a_1$ & $b_1$ & $c_1$ & $a_2$ & $b_2$ & $c_2$ \\

\hline 
12 \space & 265.85 \space & 0.8 \space & 1.5 \space & 1.45 \space & 2.5 \space &
0.19 \space & 1.89\\

\hline \\
\end{tabular}

\caption{Values of the  parameters for the pair potential.} 
\label{table1}
\end{table}

\begin{figure} 
\includegraphics[width=6.cm]{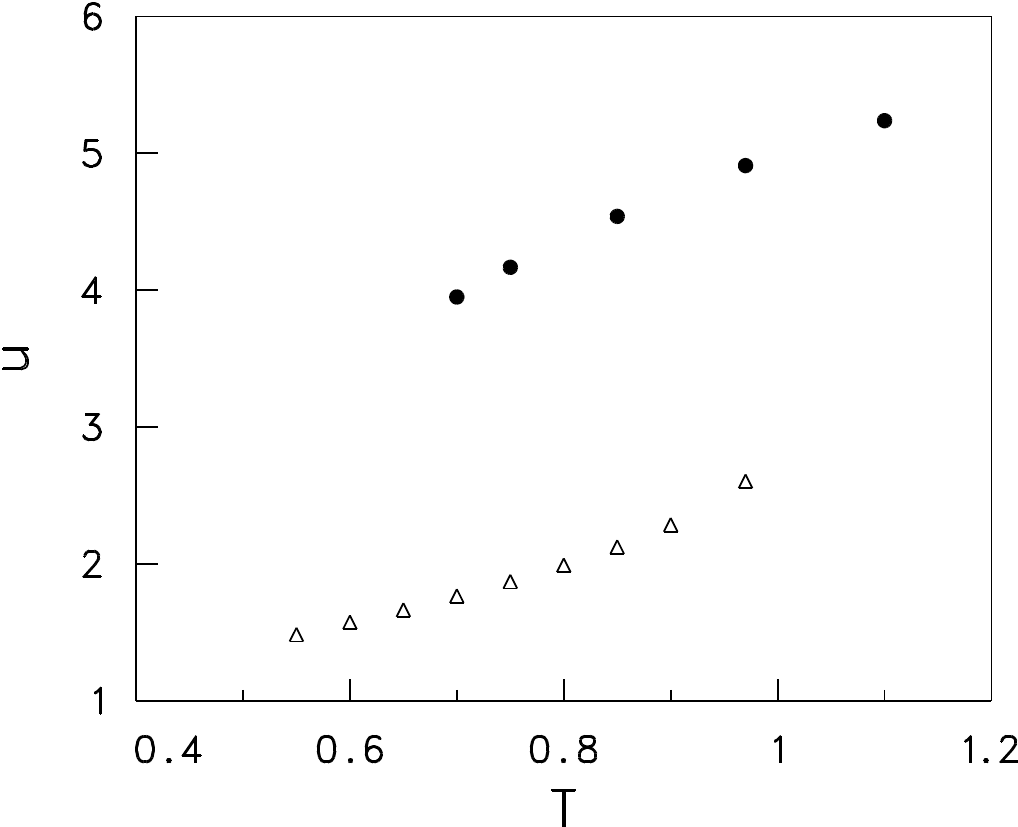}
\includegraphics[width=6.cm]{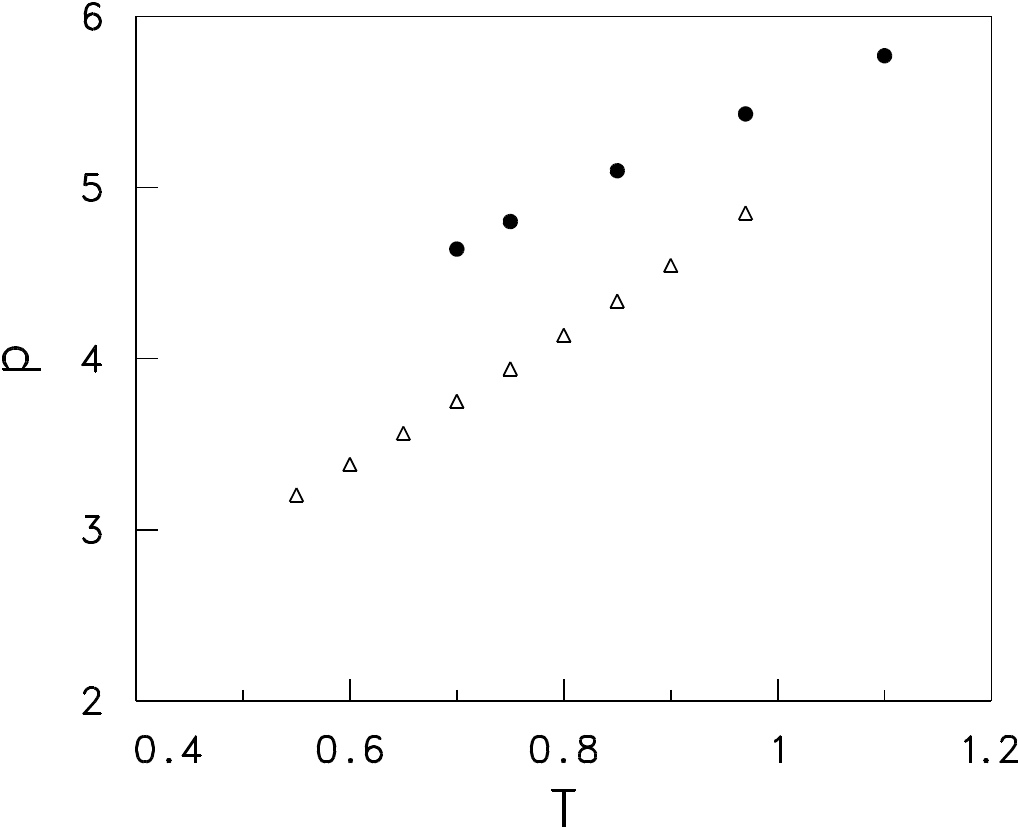}
\caption{Isochoric phase transition. Top: energy variation;
  bottom: pressure variation. Dots: high-temperature phase. Triangles: low-temperature phase. }
\label{fig2}
\end{figure}

At the beginning, the system was equilibrated in its stable isotropic
liquid state at sufficiently high temperature at the density
$\rho=0.55$. Note that this density is much below of the triple-point
density for the LJ system \cite{hansen}. We then isochorically cooled
the system, in a stepwise manner, comprehensively equilibrating it
after each temperature step. A a discontinuous change in the
parameters was detected below $T=0.65$, see Fig. \ref{fig2},
accompanied by a sharp drop in the diffusivity, an apparent signature
of a first-order phase transition to a solid phase. Accordingly, its
heating produced a significant hysteresis.  A non-trivial character of
the low-temperature phase was indicated by an anomalously long time
required for its equilibration which amounted to several billions of
time-steps.

The structure analysis of the low-temperature solid phase has been
performed by inspecting the Fourier-space pattern of its density
distribution. For that purpose, we calculated the structure factor
$S({\bf Q}) = \langle \rho({\bf Q})\rho(-{\bf Q})\rangle$, where
$\rho({\bf Q})$ is a Fourier-component of the system's number density:
\begin{equation}
\rho({\bf Q}) = \frac{1}{N} \sum_{i=1}^{N}\exp ({\bf Q r}_i) 
\end{equation} 
${\bf r}_i$ being the positions of the system's particles. $S({\bf Q})$ represents the diffraction intensity as measured in diffraction experiments. 

\begin{figure} 
\includegraphics[width=6.cm]{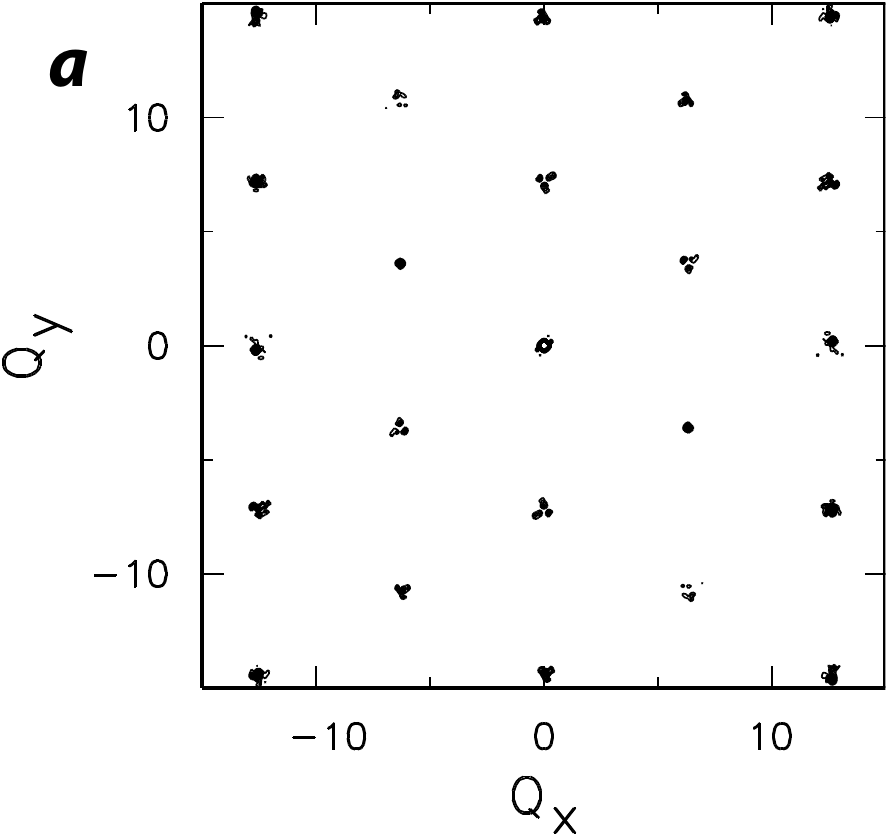}
\includegraphics[width=7.cm]{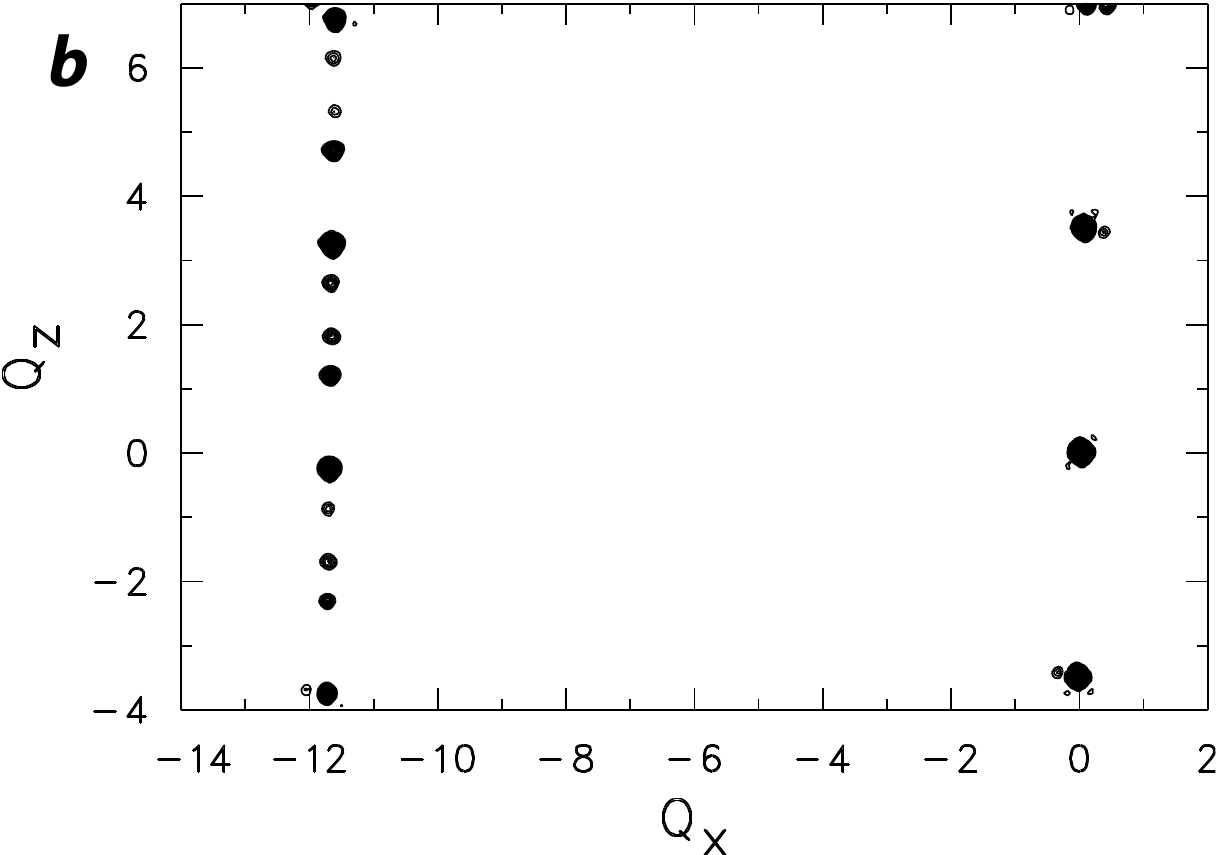}
\caption{ The isointensity plots of the structure factor $S(\bf Q)$, in two orthogonal $\bf Q$-space planes. a: $Q_z=0$;  b: $Q_y=0$. $Q_z$ denotes the axial dimension, and $Q_y$ corresponds to a translational symmetry vector, orthogonal to the axis.}
\label{fig3}
\end{figure}

As a first step, we calculated the diffraction intensity on the
$Q$-space sphere corresponding to the first peak of the spherically
averaged $S({\bf Q})$. We observed well-defined diffraction maxima
forming a regular pattern. This made it possible to determine the
global symmetry of the configuration: a single hexagonal axis was
detected. The axis orientation having been found, we calculated
$S({\bf Q})$ within two characteristic $Q$-space planes: $Q_z=0$ and
$Q_y=0$, $Q_z$ being the axis coordinate, and $Q_y$ coordinate
corresponding to a translational symmetry vector orthogonal to the
axis. The two diffraction patterns are shown in Fig. \ref{fig3}

These results compel us to make the following conclusions. First, the
observable sharpness of the diffraction maxima indicates that the
low-temperature phase produced by the phase transition is a true 3D
crystal. Moreover, the two diffraction patterns exhibit structural
features characteristic of the experimentally produced Cr-$B$ phase
\cite{leadbetter1, leadbetter2}: the configuration is a uniaxial
crystal comprised of stacked layers with dense hexagonal packing of
particles in each layer. Based on the diffraction results shown in
Fig. 3, we can also estimate the ratio of the interlayer distance to
the nearest-neighbour separation within a layer as $1.8$. 

The diffraction data presented in Fig.\ref{fig3} also provide
comprehensive information concerning the interlayer correlations in
the crystal. The hexagonal arrangement of sharp diffraction peaks in
the $Q_z=0$ plane implies the existence of global positional
interlayer correlations. The information about the type of positional
correlations of the particles of adjacent layers can be obtained by
inspecting the pattern of diffraction intensity in the axial plane,
Fig.\ref{fig3}. The hexagonal close-packed layers may be stacked with
two possible ordered arrangements: AAA..., ABA... or ABCA... where A,
B, and C denote the relative position of the layers. A random array of
ABC-type planes is also possible. These types of layer packing, or
their mixtures have been experimentally
observed\cite{leadbetter2}. The diffraction intensity profile along
the axial coordinate in Fig. 3 demonstrates four distinct auxiliary
peaks, interposed between the main peaks which represent the general
layers' periodicity. This pattern can be interpreted as representing
the predominantly ABCA-type of order in layers' packing with possible
defects \cite{leadbetter2}.  An example of this type of arrangement of
adjacent layers discerned by the real-space inspection of the
simulated  Cr-$B$ configuration is presented in Fig. 4.  \setlength{\fboxsep}{0pt}
\setlength{\fboxrule}{1pt}
\begin{figure} 

\includegraphics[height=4.1cm]{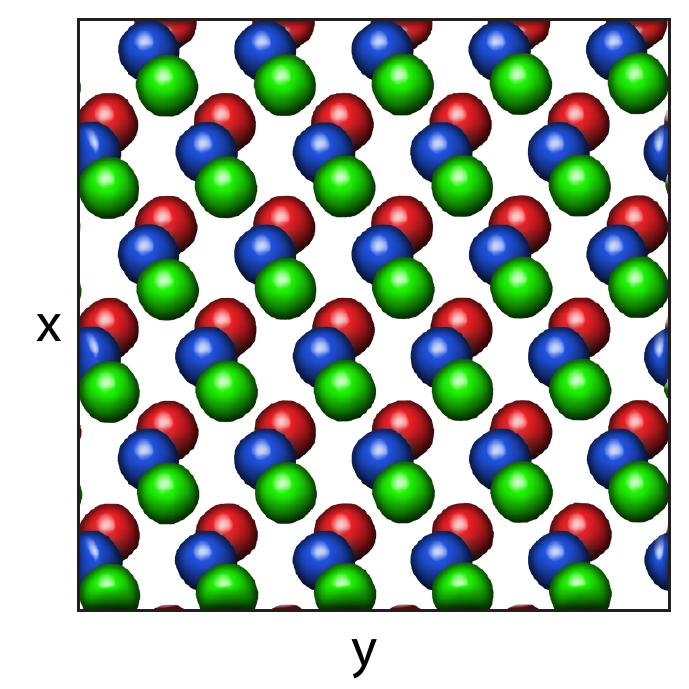}
\includegraphics[height=4.1cm]{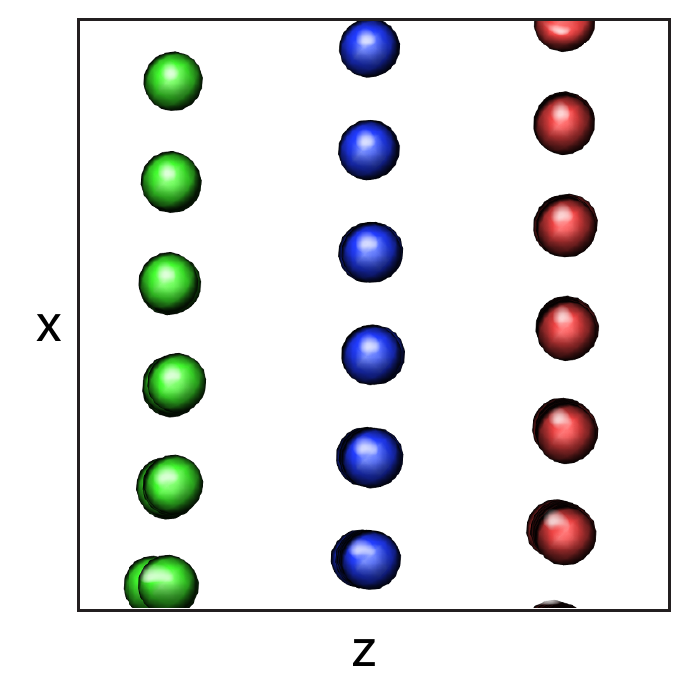}

\caption{ A fragment of the simulated Cr-$B$ configuration comprising
  three adjacent layers stacked in ABC-type sequence. Left: axial
  view; right: orthogonal view along an in-layer translational
  symmetry direction. The layers are distinguished by colors.}
\label{fig4}
\end{figure}

A subtle problem in simulation studies of the Cr-$B$ crystallisation
has always been to discriminate between the Cr-$B$ phase, possessing
true 3D long-range positional order, and the non-crystalline hexatic
phase where the hexagonal order and interlayer correlations exist only
in a limited range \cite{bates2,bates3}.  The difficulty is mainly
caused by limitations in the system size and the simulation's
time-scale which can be comparable with the space and time-scales of
the positional order in the hexatic phase. Besides, only
spherically-averaged interparticle correlations are usually considered
\cite{miguel,Tani}. In the present simulation, we were able to
identify without any ambiguity all the distinctive structural features
of the Cr-$B$ phase, including the stacking order, both in the
real-space picture and in terms of the diffraction intensity patterns.
To the best of our knowledge, this is the first reported simulation of
Cr-$B$ phase providing complete information about all the details of
its structure.

These results demonstrate that a uniaxial anisotropic structure can be
produced in a single-component system by a spherically symmetric
interparticle potential. This seemingly paradoxical result can be
rationalised by considering the structural effects of the potential's
design. This potential can be regarded as a modification of an earlier
reported pair potential \cite{dzugutov1}, judiciously designed to
induce predominantly icosahedral ordering of the first coordination
shell. It was found to produce a dodecagonal quasicrystal
\cite{dzugutov2}, and a number of other tetrahedrally ordered
structures \cite{doye}. The present potential, while retaining the
same short-range repulsive part, and the minimum position, has two
major distinctions from the earlier one. First, its minimum is much
more narrow due to a more steep attraction part. This inhibits
formation of the icosahedral ordering of the first neighbours due to
its characteristic frustration, as well as any other conceivable
densely packed structure with full first coordination shell and
energetically favours a low-density structure with a reduced number of
first neighbours.  Moreover, the extended width of the following
maximum in the present potential shifts its second repulsive part, and
thereby the second neighbours to a significantly longer distance.

As a result of this potential design, a local structure is favoured,
at appropriate density, with only six equidistant first neighbours
arranged in a hexagon. These hexagons are organised in flat densely
packed layers which are uniaxially stacked with the interlayer
distance determined by the potential's long-range repulsion.

The anisotropy of smectic phases is measured by the ratio of the
nearest-neighbour distance to the interlayer spacing. In systems of
elongated molecules, this corresponds to the degree of anisotropy
(aspect ratio) of the constituent molecules. In our system, the same
effect is induced by the potential design: the interlayer separation
is controlled by the long-range repulsion, whereas the in-layer
nearest-neighbour distance is determined by the short-range repulsion.
This implies that the apparent aspect ratio of a Cr-$B$ phase produced
in a manner we report here can be manipulated by choosing the
separation of the two repulsive branches of the potential.

At sufficiently high density, where the short-range repulsion
dominates the energy, a close-packed structure will be formed,
presumably hcp. At low densities and low temperatures, where the
structure will be determined by the long-range repulsion, the same
kind of lattice is expected to be energetically favourable. A similar
kind of isostructural polymorphism has been reported for a stepwise
pair potential \cite{Bolhuis2}.

We conclude with the following remarks. 

1. So far, colloidal smectic phases have only been found to appear in
systems of rod-like colloidal particles \cite{maeda}. The
spherically-symmetric nature of the interparticle interaction in the
present model, and the similarity of its main features to classical
DLVO theory for colloidal interaction \cite{DL,VO,ims-forces} (amended
with hard core repulsion or steric repulsion at close to contact),
suggests a possibility that smectic-like layered structures can be
produced in colloidal systems of spherically shaped particles, with
appropriate tuning of the effective force field.

2. The local structural isomorphism of the Cr-$B$ and fluid hexatic
phase suggests that the latter too can possibly be produced using the
general approach to the potential design we exploited here, as well as
the smectic $A$ phase.

3. An intriguing anomaly of the Cr-$B$ dynamics is the presence of
soft shear modes tentatively concluded from spectroscopic measurements
\cite{fera}. The ultimate simplicity of the present model can
facilitate investigation of the underlying mechanism of this dynamical
singularity.

In summary, we used a molecular-dynamics simulation to demonstrate
that the  Cr-$B$ phase that has so far only been observed in
mesogenic systems composed of anisotropic molecules can be formed in a
system of identical particles interacting via a spherically-symmetric
potential. This finding remarkably simplifies the basic model of
smectic phases, thereby advancing our understanding of the causes
underlying the occurrence of particular structures in the phase
transformations of liquid crystals.

We gratefully acknowledge the valuable assistance of Michael
Schliepake and other staff members of the Centre for Parallel
Computers (PDC), KTH. Document release number LLNL-JRNL-639781;
prepared by LLNL under Contract DE-AC52-07NA27344.

\end{document}